\documentclass[useAMS,usenatbib]{mn2e}
\usepackage{graphicx}
\title[NGC~3310 and Its Tidal Debris:  Remnants of Galaxy Evolution]{NGC~3310 and Its Tidal Debris:  Remnants of Galaxy Evolution}
\author[E. H. Wehner, J. S. Gallagher, P. Papaderos, U. Fritze-von Alvensleben and K. B. Westfall]
{E. H. Wehner$^{1}$\thanks{E-mail: wehnere@physics.mcmaster.ca}, J. S. Gallagher$^{2}$, P. Papaderos$^{3}$, U. Fritze-von Alvensleben$^{4}$, \newauthor and K. B. Westfall$^{2}$  \\
$^{1}$Department of Physics and Astronomy, McMaster University, 1280 Main Street West, Hamilton, ON L8S 4M1, Canada \\
$^{2}$Department of Astronomy, University of Wisconsin-Madison, 475 North Charter Street, Madison, WI  53717, USA \\
$^{3}$Universit\"ats-Sternwarte, Institut f\"ur Astrophysik, Friedrich-Hund Platz 1, 37077 G\"ottingen, Germany \\
$^{4}$University of Hertfordshire, Centre for Astrophysics Research, College Lane, Hatfield AL10 9AB, UK 
}

\begin{document}

\date{Accepted 2006 June 27. Received 2006 June 27; in original form 2006 April 20}

\pagerange{\pageref{firstpage}--\pageref{lastpage}} \pubyear{2006}

\maketitle

\label{firstpage}

\begin{abstract}

NGC~3310 is a local galaxy with an intense, ongoing starburst thought to result from a merger with a companion galaxy.  It has several known tidal features in the northwest and southern regions around the main galactic disc, as well as a closed, tidal loop emerging from the eastern side of the disc and rejoining in the north.  This loop appears to be distinct from the rest of the shells surrounding NGC~3310 and is the first of its kind to be detected in a starburst galaxy.  In this work, we present $UBVR$ photometry to faint surface brightness levels of this debris network, and we explore various strategies for modelling NGC~3310's disc and subtracting its contribution from each region of debris.  We then compare these photometric results with the GALEV spectral synthesis models, and find possible material from the intruder galaxy, suggesting that the recent accretion of several small galaxies is driving the evolution of NGC~3310.

\end{abstract}


\section{Introduction}
\label{secintro}

NGC~3310 is a nearby disc galaxy undergoing an intense starburst.  This system is
particularly intriguing because it is a field galaxy which displays signs of a
recent and complex merging history.  At a distance of about 14 Mpc ($H_0 = 72$
km s$^{-1}$ Mpc$^{-1}$) and with a radial extent of 16 Kpc, including its most far-flung stellar debris, 
this relatively small galaxy is a system in turmoil.
It has an HI mass of $2.2 \times 10^{9} M_{\odot}$, an estimated total mass of
$2.2 \times 10^{10} M_{\odot}$ \citep{ks01} and a global star formation rate 
of 8.5 M$_{\odot}$ yr$^{-1}$, most of which is
occurring in a starburst ring surrounding the nucleus.  There are also many star
forming regions throughout the disc, as well as young star clusters which
\citet{deGrijs03a}
find to be approximately 40 Myr old.  This current starburst is undoubtedly
largely responsible for the galaxy's blue global colour of $B-V = 0.32$ 
\citep{deVauc91}.  As the light from this galaxy is dominated by young stars, 
separating the contributions of the starburst and the underlying disc to the overall
colour of NGC~3310 is difficult.

There is much evidence that NGC~3310 has recently undergone at least one
significant merger.  This idea was first proposed by \citet{bh81}, who
suggested that a dwarf galaxy collided with NGC~3310 and that the famous ``arrow" on
the western side consisted of the remains of the interloper.  In addition to the
visible optical debris, such as numerous shell-like features and the ``bow
and arrow" \citep{wc67}, the radial surface brightness profiles of NGC~3310 were found to
follow an $R^{1/4}$ law, which has been interpreted by \citet{smith96} 
as evidence that the main disc has been disturbed by a recent merger.  
Furthermore, while the nucleus has almost solar metallicity, the disc is substantially more
metal poor \citep{hb80,phm90,pastoriza93}. Because of this discrepancy, it has been suggested (see \citet{deGrijs03a}
for a discussion) that NGC 3310 collided with a gas rich, metal-poor companion galaxy.  Another 
possibility, given the low metallicity of the disc, is that this galaxy is in the process
of forming a new stellar disc.

Work by \citet{ks01} suggests that the arrow feature is not a distinct
kinematic element, but rather
the beginning of an extended HI tail which is found in the velocity range of
$804-953$ km/s.   When this tail is considered in conjunction with the Southern HI tail,
the total mass of the interloper approaches almost 10 per cent of the galaxy's total
mass.  In \citet{wehner05} (hereafter Paper I), we announced the discovery
of an additional arc of stellar debris.  This arc is the first closed stellar loop found in
a starburst galaxy, and its location does not coincide with the gaseous, $HI$ tidal tails.  
With a total stellar mass of $4 \times 10^8 M_{\odot}$ this
large loop supports the idea that NGC~3310 underwent a larger merger than
previously thought.  The possibility also remains that NGC~3310 suffered numerous small
collisions, rather than one large merger.

There are several ways to further explore the merging history of this galaxy.  In
Paper I we defined photometric regions, and in this work we continue our study of
them.  Our goal is to explore the origins of these debris features and to determine whether
they are shell-like structures created by stars pulled from NGC~3310's own disc, as
in the weak interaction model of shell formation \citep{tw90}, or whether they represent the
remains of another galactic companion.  We then can use this information to
gain insight into how the merger(s) may have progressed.  In \S~\ref{secdata}, we present
$UBVR$ photometry for all previously defined regions of debris.  We
discuss the significance of this debris network's photometric colours and use several
different methods to compare them to the colours of NGC~3310's underlying disc. 
We explore disc modelling strategies for this disturbed
galaxy, and present these results in \S~\ref{secphotom}.   In \S~\ref{secgalev}, we use spectral
synthesis models to further identify the star formation histories contained in the
debris regions themselves and compare this to the current ideas of the star formation
history of the main disc.  In \S~\ref{secsummary}, we then discuss the consequences of these
results for the merging history of NGC~3310.

\section{Observations and Data Reduction}
\label{secdata}

\subsection{Broad-band Imaging}
\label{subsecimaging}

For this study we make use of data from both the WIYN 3.5m\footnote{The WIYN 
Observatory is a joint facility of the University of Wisconsin--Madison, Indiana 
University, Yale University, and the National Optical Astronomy Observatory.}
and the WIYN 0.9m
telescopes.  The reductions for the WIYN 0.9m data, taken on the nights of
2003 March 7-11, are described in Paper I.  The WIYN 3.5m data were
taken with the MiniMosaic CCD camera on the night of 2004 April 20, under
photometric conditions.  For each night, numerous Landolt standard fields were
observed in order to obtain photometric solutions.  While a range in airmass
was covered for all 0.9m data, standards were only taken at the WIYN 3.5m at
the beginning and the end of the night.   Thus, we used the standard extinction
coefficients for Kitt Peak and solved only for the colour terms in our data.

The MiniMosaic data were reduced using both IRAF\footnote{IRAF is distributed 
by the National Optical Astronomy Observatories, which are operated by the 
Association of Universities for Research in Astronomy, Inc., under cooperative 
agreement with the National Science Foundation.} and MIDAS\footnote{Munich 
Image Data Analysis System, provided by the European Southern Observatory 
(ESO).}.  In IRAF, the MSCRED package was used to perform the overscan
correction, which was done on a line by line basis for these data.
It was also used to create the average bias and dome flat frames and
for the final bias subtraction and flat division.  The two separate
chips of the WIYN 3.5 MiniMosaic CCD Camera were then combined into 
a single image with the MSCIMAGE
task, which remaps the two chips onto a single coordinate system.
Once the two chips were merged, the multiple exposures of NGC~3310
were combined with imcombine and bad pixel masks were used to remove the gap
left from the separate mosaic chips.  

\subsection{H$\alpha$ Images}
\label{subsechalpha}

The H$\alpha$ images of NGC~3310 were obtained on the night of April
7, 2003, using the WIYN 3.5m telescope under photometric conditions with seeing of
$\sim 1\arcsec FWHM$.  The images were processed in the manner
described above before the final H$\alpha$ image was created.  The
variable seeing throughout the night led to a small difference in
the FWHMs between the broad and narrow-band data.  The FWHM of the H$\alpha$ image
was approximately 1.1 pixels, or 0.16\arcsec,  larger than that of the $R$-band image.  
To account for this, we used
PSFMATCH to convolve the R-band data with a Gaussian to broaden the
FWHM profile, while still conserving the total flux.  We then
calculated the flux-ratio for stars surrounding the galaxy in the narrow and broadband
data, scaled the R-band down accordingly and subtracted it from the H$\alpha$.
This effectively removed the continuum from the H$\alpha$ image and
as a result, removed most of the stars.  

This final image can be
seen in Fig. 1.  The regions of interest defined in Paper I and the SparsePak 
fibre configuration have been overlaid on 
this H$\alpha$ image.  Since H$\alpha$ emission is associated with stellar complexes having ages of $<$10~Myr, Fig. 1 demonstrates how active star formation has retreated to the inner parts of NGC~3310, and is not progressing at significant levels in most of the outer galaxy.

%
%

\begin{figure*}
\begin{minipage}[t]{160mm}
\begin{center}
\begin{tabular}{cc}
\includegraphics[width=3.15in]{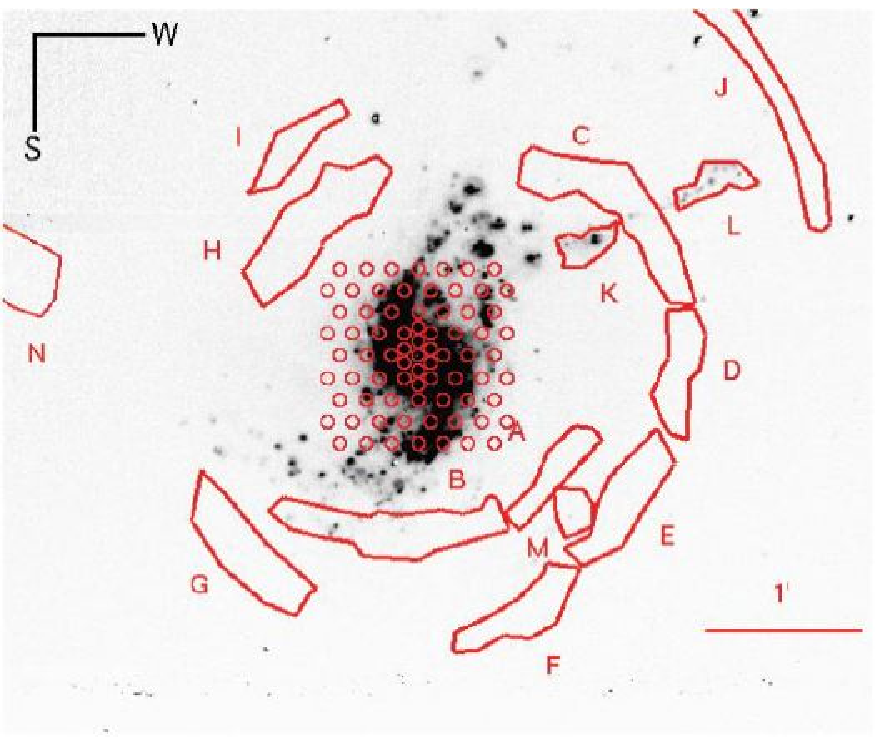} &
\includegraphics[width=3.15in]{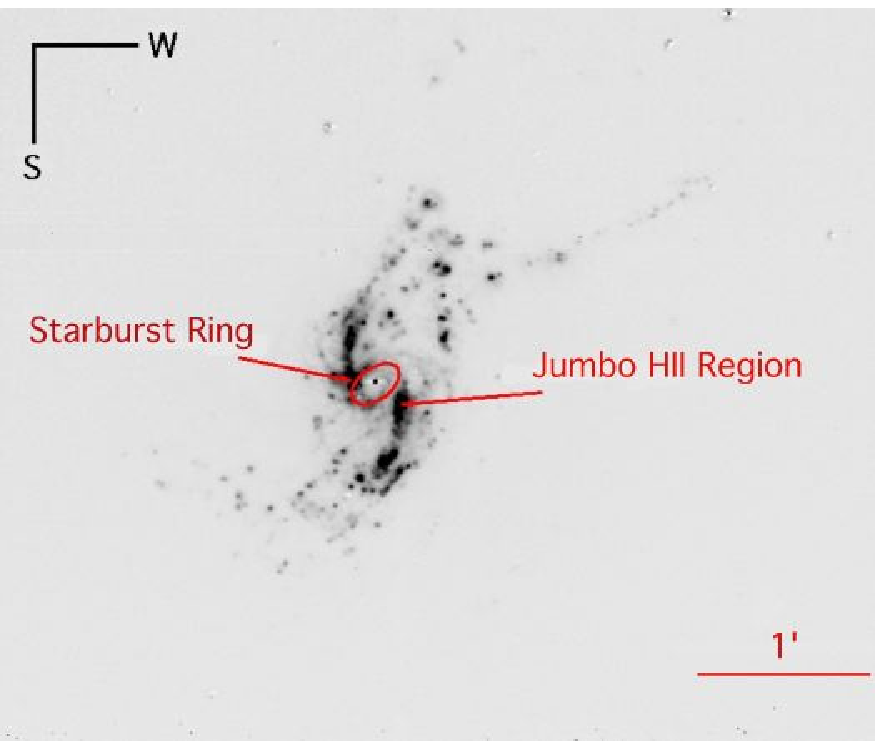}
\end{tabular}
\caption{Continuum subtracted $H\alpha$ image of NGC~3310 taken with the WIYN 3.5m.  Regions 
of debris defined in Paper I have been overlaid in Figure 1a (left), as well as the SparsePak array of fibres.  From this image, we can determine that
there is little to no star formation in all regions of debris, except for those which
define the ``arrow" (regions K \& L).  Figure 1b (right) shows more of the structure in the H$\alpha$ emission, and the well known Jumbo HII region and the circumnuclear starburst ring are labelled. }
\label{f1}
\end{center}
\end{minipage}
\end{figure*}

\subsection{Spectroscopy}
\label{subsecspec}

The SparsePak \citep{bershady05, bershady04} data for NGC~3310 were obtained on the WIYN 3.5m in
Kitt Peak, Az, on December 14, 2005.  Our observations were made under
photometric conditions with optimal seeing (0.5\arcsec).  We used the
echelle 316@63.4 grating at order 8, centred at $\lambda = 6655\AA$, resulting in a
resolution of 11,400 or 0.21 $\AA$ per pixel.  This corresponds to a resolution
of 20 km s$^{-1}$ over a two pixel resolving element.
The fibre array was centred on the galaxy's bulge, and the fibre positions are
shown in Fig. 1.  We obtained 2 exposures of 1800s each.
We also took calibration arcs using the CuAr lamp throughout the
night.

The basic reductions, such as the bias and overscan corrections, for 
these data were done using IRAF with the CCDRED and HYDRA packages.  
The combined images and calibration frames were created using the IMCOMBINE
procedure and subsequent was completed using the DOHYDRA task.
Using this task, we flat-fielded the spectra and performed the wavelength calibration with
the assistance of the NOAO calibration spectra 
database\footnote{http://www.noao.edu/kpno/specatlas/index.html}.  The
spectra were linearised to facilitate line extraction and further
analysis.  Also, we used the SparsePak sky fibre data to remove the
sky lines from the final spectra.  Of the seven sky fibres, four
were found to contain target H$\alpha$ lines and were removed.  The
remaining three fibres were averaged and used for sky subtraction.

\section{Photometry}
\label{secphotom}

\subsection{Integrated Colours}
\label{subsecintcolours}

The colours for each region are listed in Table 1, and the $UBVR$ 
surface photometry for the network of debris in NGC~3310 is shown in
Table 2.  These calibrated magnitudes and their errors were calculated using the POLYPHOT package in IRAF
and are described in Paper I.  The errors shown include only random sources of error, and thus
represent lower limits.  However, for low surface brightness photometry, the main source of
systematic error is flat field error.  With the S2KB chip and the well-baffled WIYN 0.9m
telescope, the signal is flat across the chip to within 0.7 percent.  
Thus, the variation across each region shown in Fig. 1 is negligible for the 0.9m data.  
However, for the $U$-band data, taken on the WIYN 3.5m, scattered light becomes a more serious
issue and the field is less flat by an order of magnitude.   As a result, the errors in $U-B$ 
may be as high as $\pm 0.1$ mag.  Another source of systematic error is that
the natural variation in the colour within each region.  However, average
colours are still useful for studying trends as well as the aggregate properties of the stellar
populations within each region.  

While there is quite a range in $B-V$ colour
across the system, in general, most of the defined regions have bluer colours than
are typical of non-starbursting disc galaxies \citep{deJong96}.   Ideally, we 
would like to compare the debris colours
with those of NGC~3310's underlying disc.  The difficulty with this comes from the
fact that when we observe NGC~3310, we are seeing light from both
the underlying disc and the overlying starburst, combined.  

\begin{table*}
    \begin{minipage}[t]{140mm}
	\begin{center}
    	\caption{Colour Indices of the Tidal Debris of NGC~3310}
\label{tab1}

	\begin{tabular}{@{}lccccccc}
  	  \hline
	  Region & $U$ & $B$ & $V$ & $R$ & $U-B$ & $B-V$ &  $V-R$ \\
	  \hline

   A & 17.43$\pm$0.006 & 17.49$\pm$0.02 & 17.08$\pm$0.003 & 16.73$\pm$0.004 & -0.06$\pm$0.02 & 0.41$\pm$0.02 & 0.36$\pm$0.007 \\ 
   B & 16.23$\pm$0.003 & 16.41$\pm$0.01 & 15.93$\pm$0.002 & 15.53$\pm$0.002 & -0.18$\pm$0.01 & 0.48$\pm$0.01 & 0.40$\pm$0.004 \\ 
   C & 16.01$\pm$0.002 & 16.20$\pm$0.01 & 15.77$\pm$0.002 & 15.39$\pm$0.002 & -0.18$\pm$0.01 & 0.42$\pm$0.01 & 0.39$\pm$0.004 \\ 
   D & 17.21$\pm$0.006 & 17.55$\pm$0.02 & 17.12$\pm$0.004 & 16.73$\pm$0.005 & -0.34$\pm$0.03 & 0.43$\pm$0.03 & 0.39$\pm$0.009 \\ 
   E & 17.08$\pm$0.006 & 17.41$\pm$0.03 & 16.89$\pm$0.004 & 16.49$\pm$0.005 & -0.33$\pm$0.03 & 0.52$\pm$0.03 & 0.41$\pm$0.009 \\ 
   F & 17.35$\pm$0.007 & 17.64$\pm$0.03 & 17.01$\pm$0.004 & 16.59$\pm$0.005 & -0.29$\pm$0.03 & 0.63$\pm$0.03 & 0.42$\pm$0.009 \\ 
   G & 17.46$\pm$0.009 & 18.16$\pm$0.05 & 17.63$\pm$0.013 & 17.24$\pm$0.010 & -0.70$\pm$0.06 & 0.53$\pm$0.06 & 0.39$\pm$0.017 \\ 
   H & 16.54$\pm$0.004 & 16.74$\pm$0.02 & 16.31$\pm$0.003 & 15.93$\pm$0.003 & -0.20$\pm$0.02 & 0.43$\pm$0.02 & 0.38$\pm$0.006 \\ 
   I & 17.98$\pm$0.010 & 19.26$\pm$0.10 & 18.92$\pm$0.025 & 18.65$\pm$0.023 & -1.28$\pm$0.10 & 0.34$\pm$0.10 & 0.27$\pm$0.038 \\ 
   J & 16.22$\pm$0.004 & 17.83$\pm$0.05 & 17.22$\pm$0.008 & 16.72$\pm$0.007 & -1.61$\pm$0.05 & 0.61$\pm$0.05 & 0.50$\pm$0.013 \\ 
   K & 16.89$\pm$0.003 & 17.48$\pm$0.02 & 17.27$\pm$0.003 & 16.91$\pm$0.004 & -0.59$\pm$0.02 & 0.22$\pm$0.02 & 0.36$\pm$0.006 \\ 
   L & 17.17$\pm$0.004 & 18.05$\pm$0.03 & 17.79$\pm$0.007 & 17.39$\pm$0.006 & -0.88$\pm$0.03 & 0.26$\pm$0.03 & 0.41$\pm$0.011 \\ 
   M & 18.00$\pm$0.007 & 18.18$\pm$0.03 & 17.77$\pm$0.004 & 17.44$\pm$0.006 & -0.19$\pm$0.03 & 0.41$\pm$0.03 & 0.34$\pm$0.010 \\ 

	  \hline
	\end{tabular}
	\end{center}
    \end{minipage}
\end{table*}


While modelling these two
components separately is tricky to do, one can use several different approaches
to estimate the colour of the underlying disc.  If we assume that
before the collision and its subsequent starburst, NGC~3310 was a typical disc
galaxy, we can use its RC3 \citep{deVauc91} classification ($T = 4.0\pm0.3$) to compare its colour with 
the average colour for similar galaxies.  According to the work of \citet{deJong96}
galaxies of a similar classification type to NGC~3310 have a $B-V = 0.74$.
Each tidal feature in NGC~3310 is significantly bluer than this.
Since the weak interaction model suggests that shells are formed from a galaxy's
own disc stars, we expect that shells formed in this way would have stellar
populations similar to the galaxy's main disc.  Thus, their blue colour suggests
it is unlikely that NGC~3310's debris was formed by a weak interaction.  However,
this assumption breaks down if the merger(s) that NGC~3310 experienced caused it to
change its structural class or if NGC~3310's underlying disc, due to its starburst 
and possible recent interactions, is simply not representative of an ordinary spiral galaxy, 
and thus its RC3 classification isn't as well correlated to its current colour as a normal galaxy would be.  As discussed in further sections, this latter argument appears to be the case.

\begin{table*}
  \centering
  \begin{minipage}[t]{140mm}
    \caption{Photometric and Physical Properties of the Tidal Debris of NGC~3310}

    \label{tab2}
    \begin{tabular}{@{}lccccccccccc}
 	\hline
	Region & $\mu_U$ & $\mu_B$ &  $(S/N)_{pix}$\footnote{Signal-to-noise per pixel on the final calibrated image.}  & $(S/N)_{ap}$\footnote{Total signal-to-noise for each aperture.} & $\mu_V$ & $(S/N)_{pix}$ & $(S/N)_{ap}$ & $\mu_R$ & $(S/N)_{pix}$ & $(S/N)_{ap}$  \\
	\hline
   	A & 24.10 & 24.16 & 0.28 & 10.09 & 23.75 & 1.52 & 54.64 & 23.39 & 2.13 & 76.47 \\ 
   	B & 23.94 & 24.12 & 0.29 & 16.90 & 23.64 & 1.68 & 97.41 & 23.24 & 2.44 & 142.01 \\ 
   	C & 23.80 & 23.99 & 0.33 & 19.80 & 23.56 & 1.80 & 108.52 & 23.18 & 2.59 & 155.92 \\ 
   	D & 24.17 & 24.51 & 0.20 & 8.37 & 24.09 & 1.12 & 45.95 & 23.69 & 1.62 & 66.63 \\ 
   	E & 24.38 & 24.71 & 0.17 & 8.15 & 24.19 & 1.01 & 48.61 & 23.79 & 1.48 & 71.28 \\ 
   	F & 24.40 & 24.70 & 0.17 & 7.41 & 24.07 & 1.13 & 48.71 & 23.65 & 1.69 & 72.38 \\ 
   	G & 24.98 & 25.67 & 0.07 & 3.71 & 25.15 & 0.42 & 22.37 & 24.76 & 0.61 & 32.32 \\ 
   	H & 24.32 & 24.52 & 0.20 & 12.12 & 24.09 & 1.11 & 66.76 & 23.71 & 1.59 & 95.50 \\ 
   	I & 24.55 & 25.83 & 0.06 & 2.06 & 25.50 & 0.31 & 10.51 & 25.22 & 0.40 & 13.65 \\ 
   	J & 24.11 & 25.72 & 0.07 & 4.26 & 25.11 & 0.43 & 27.41 & 24.61 & 0.69 & 43.74 \\ 
   	K & 22.79 & 23.38 & 0.57 & 14.37 & 23.16 & 2.60 & 65.42 & 22.80 & 3.65 & 91.84 \\ 
   	L & 23.33 & 24.21 & 0.27 & 7.59 & 23.95 & 1.26 & 35.88 & 23.54 & 1.85 & 52.57 \\ 
   	M & 23.88 & 24.06 & 0.31 & 7.66 & 23.65 & 1.67 & 41.61 & 23.31 & 2.29 & 57.12 \\ 
   	N & -- & -- & 0.00 & 0.00 & 25.55 & 0.29 & 13.34 & 25.44 & 0.33 & 14.95 \\ 
	\hline
    \end{tabular}
  \end{minipage}

\end{table*}

\subsection{Disc Models}
\label{subsecdiscmodels}

We turn to disc modelling to further explore the properties of the underlying
disc.  For this task, we used several different MIDAS scripts to calculate the
surface brightness as a function of radius.  For this we used the FIT/ELL3
routine in MIDAS, which is based on the iterative algorithm of
\citet{bh87}.  Thus, the resulting surface brightnesses represent the two-dimensional
flux distribution based on a $\chi^2$ convergence fit.  As such, these
profiles include light from the underlying disc, the
starburst, and overlapping stellar debris.  Ideally we would like to
separate these components in order to compare the colours of the stellar
debris with that of the underlying disc of NGC~3310.

%
%
\begin{figure*}
\begin{minipage}[t]{160mm}
\begin{center}
\begin{tabular}{ll}
\includegraphics{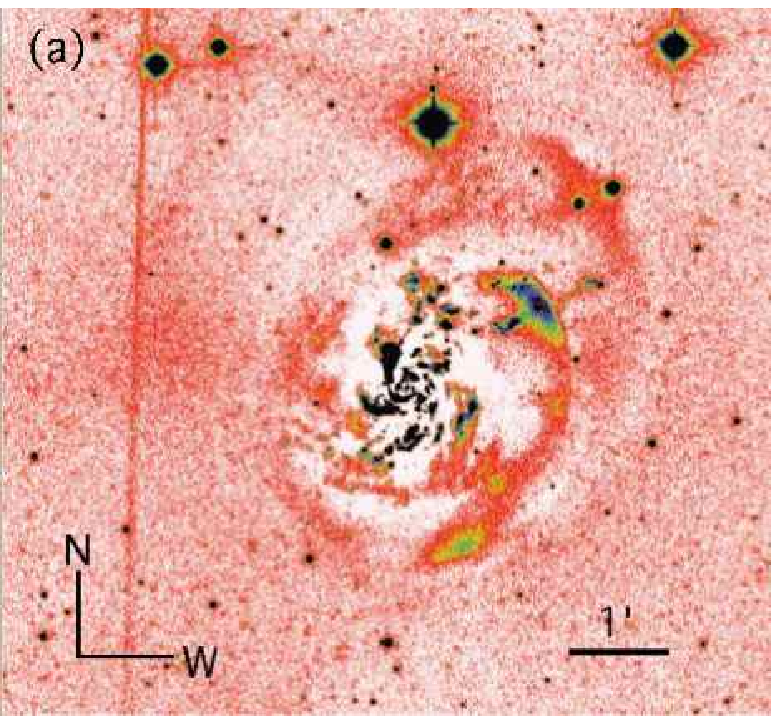} &
\includegraphics{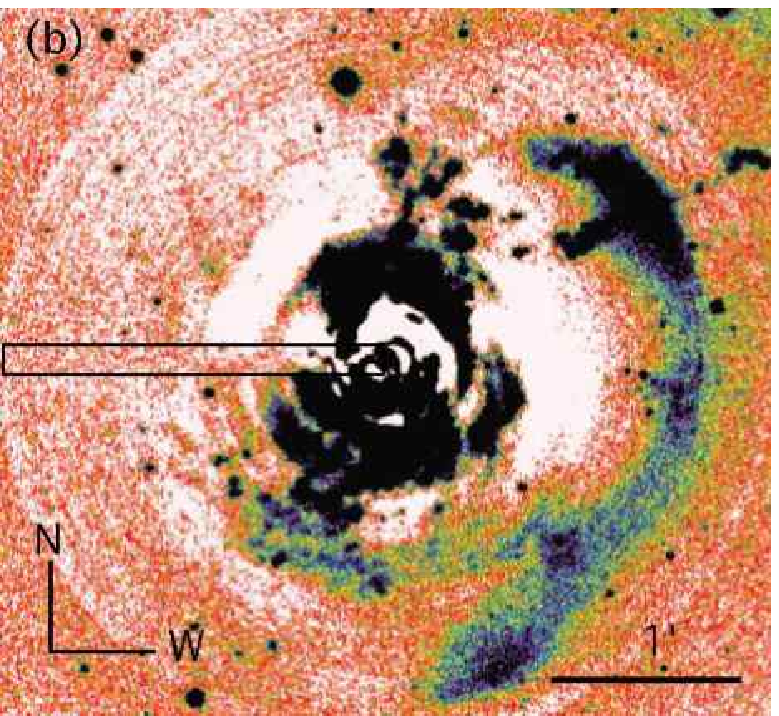} \\
\includegraphics{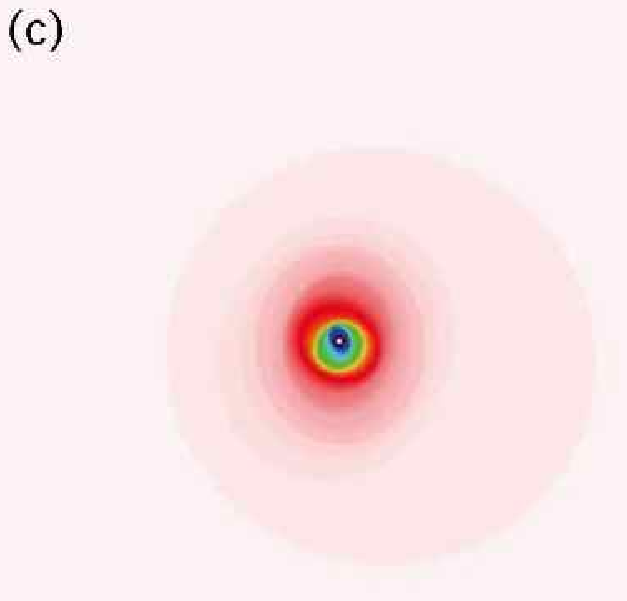} & 
\includegraphics{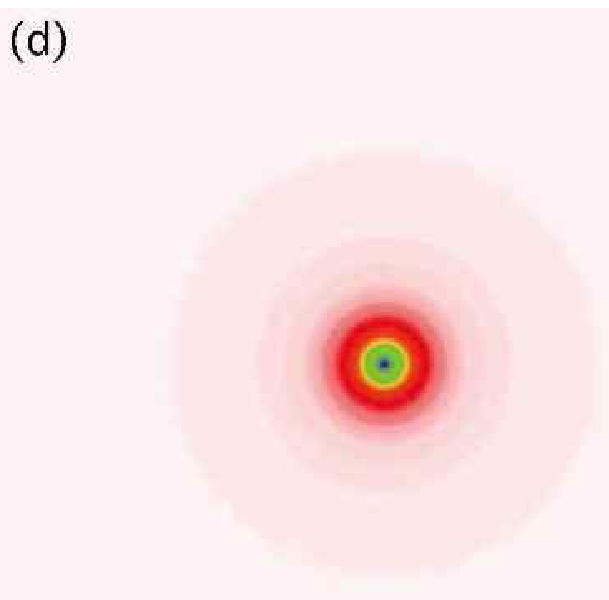} \\
\end{tabular}
\caption{\emph{Left:} \textbf{(a)} Residuals obtained by subtracting the 2D flux distribution \textbf{(c)} calculated from the FIT/ELL3 script in MIDAS from the $B$-band image.  \emph{Right:}  \textbf{(b)} Residuals obtained by subtracting the circularly symmetric 2D light distribution \textbf{(d)} from the $B$-band image.  The black rectangle in \textbf{(b)} indicates the slice of the Eastern side of the galaxy used to create the circular model.}
\label{f2}
\end{center}
\end{minipage} 
\end{figure*}

%
%
\begin{figure}
\begin{center}
\begin{minipage}[t]{3.3in}
\begin{tabular}{c}
\includegraphics[width=3.15in]{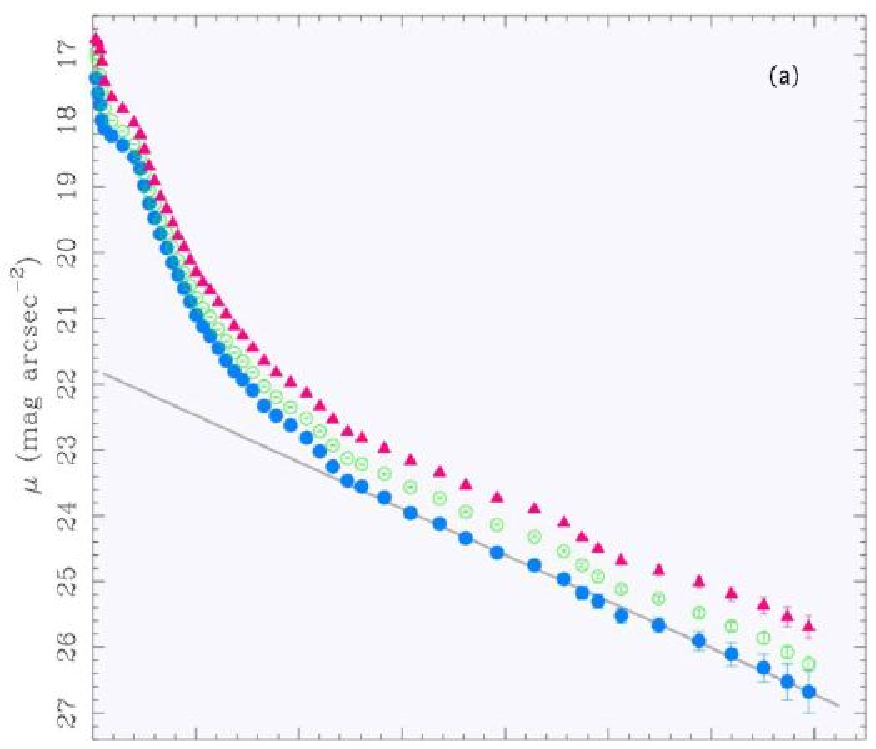} \\
\includegraphics[width=3.15in]{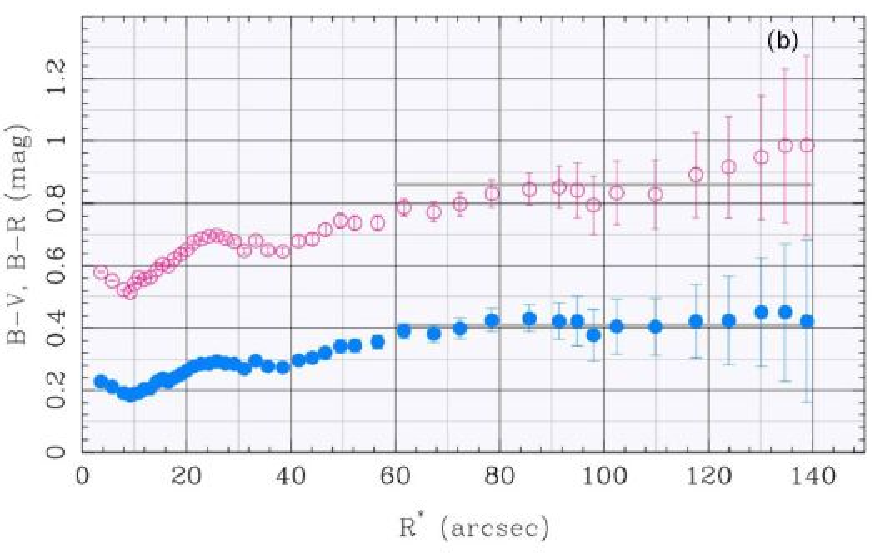} \\
\end{tabular}
\caption{ \textbf{(a)} Surface brightness profiles in $BVR$ for NGC~3310.  The grey line represents a fit to the $B$-band profile at radii $\ge$ 60$\arcsec$.  The $R$-band is represented by red triangles, $V$ by green open circles and $B$ by blue closed circles.  When error bars are not shown, the errors do not exceed the size of the point.  \textbf{(b)} Radial colour profiles for NGC~3310.  The $B-V$ profile is shown in blue, closed circles, and the $B-R$ profile is shown with red, open circles.}
\label{f3}
\end{minipage}
\end{center}
\end{figure}

Once we obtained our best 2D flux distribution for NGC~3310, we subtracted
it from our original image in order to examine the structure of the
debris.  This image is shown in Fig. 2a.  In this image, the debris
network and the clumpy star-forming regions are clearly visible.  However,
when we examine the resulting 2D distribution, we find that the presence of
debris distorts the fit on the western side (Fig. 2c)
In order to correct for this, we wrote a new MIDAS script to use a slice of the 
eastern portion of the galaxy, shown as a black rectangle in Fig. 2b, which we
then azimuthally expanded to represent the entire surface of the disc.
This is only a partial solution in that it still
contains light from some star-forming regions, but it allows us to sample the less
optically disturbed side of the galaxy and to choose a slice containing no bright tidal
debris.  We then subtracted this fit from our image, shown in Fig. 2b.
The actual 2D fit from this circular model is shown in Fig. 2d.
In comparing Figs. 2a and 2b, one can clearly see the asymmetries induced by 
the western regions of tidal debris in
the outer disc in 2a.  These features disappear when using the eastern
slice to model the disc.

Next, we applied our photometric solutions to calibrate the model frames in each band, and subtracted
them to create a $B-V$ frame for the disc model.  We then used the $D_{25} = 3\arcmin.1$ value
defined in RC3 as our diameter and performed aperture photometry to find a total $B-V$ of
0.30.  This agrees with previous measurements of the
global $B-V$ for this system and provides a useful consistency check on our
2D flux distributions.  While these models do not
separate the disc and starburst contributions to the total emission,
they are effective at subtracting the outer regions of the disc,
since at higher radii the contribution from the starburst becomes
insignificant.  The residuals from this model, which highlight clumpy galactic regions 
and regions of debris, can be seen in Fig. 2b.

In order to carry out detailed surface brightness photometry of NGC~3310 we then
derived the $BVR$ profiles using method iv of \citet{papad02}.  These results are
shown in Figs. 3 and 4.  In Fig. 3a, the surface brightness is shown as a
function of radius plotted linearly.  The solid line is a fit to the $B$-band
surface brightness profile (SBP) for $R \ge 60\arcsec$, and yields an approximation for the
disc and debris at these radii.  This fit implies a central surface brightness of 
$\mu_0$=21.77 $\pm$ 0.06 mag arcsec$^{-2}$, comparable to the Freeman value of 
21.65~mag~arcsec$^{-2}$ \citep{freeman}. This value is somewhat lower than that found
by \citet{bba98} ($\mu_V = 23.98$) but our models include contributions from both
the bulge and the disc, and can therefore be expected to be brighter.
We also find an exponential scale length of 2.08$\pm$0.04 Kpc, or $31\arcsec$, 
and an absolute total B magnitude of -18.4 mag (for D $=14$ Mpc) for the disc.  This 
scale length is somewhat longer than that of \citet{duric96} ($l = 7.3\arcsec$) for
an off-H$\alpha$ waveband and significantly smaller than the $l = 123\arcsec$ measured
by \citet{bba98} in the $V$-band.  However, the SBPs we present in Figs. 3 and 4 extend at least
two magnitudes deeper than those previously measured.

Fig. 3b shows the $B-V$ and $B-R$ radial colour profiles for NGC~3310, calculated by
subtracting the respective SBPs.  The colours for the outer
disc region (defined as $R \ge 60\arcsec$) are $B-V =$ 0.42 $\pm$ 0.03 and $B-R = $
0.86 $\pm$ 0.07 magnitudes.  

It is also interesting the shape of the surface brightness profiles for each band.  
In Fig. 4 we plotted the SBPs against R$^{1/4}$ in order to check to how well the SBP 
can be approximated by the de Vaucouleurs law.  From this plot we see that the de 
Vaucouleurs law (fit by $r^{1/\eta}$ where $\eta = 4$) describes parts of the SBP (between R$^{1/4}=1.7$ and R$^{1/4}=2.7$) 
very well, but not the SBP as a whole.  A S\'ersic fit yields an $\eta = 2.8$ and a slightly better 
$\chi^2$ than a model with a fixed $\eta$ = 4.  However, all S\'ersic distributions leave 
systematic residuals when fitted to the total SBP, indicating that a single fitting 
law cannot perfectly fit the SBP.  In fact, an exponential disc appears to emerge at higher radii.

It is clear from Figs. 3 and 4 that the light from the starburst makes a large contribution
to the overall light from NGC~3310.  In order to better understand the nature of 
NGC~3310's underlying disc, we must find
some way to eliminate the light contribution to the bright starbursting regions.
To accomplish this, we have created pixel masks for each of the $B$, $V$ and $R$-band
images in order to remove the contribution of obvious star-forming regions across
the disc.  The same mask was used for all images, and was created by eliminating all counts
greater than 5$\sigma$ above the sky in the B-band image, which most closely represents 
the younger star-forming regions.  While masking reduces the total number of 
pixels available to the
models, we have sufficient pixels in regions outside the main starburst ring
to measure colours with a reasonable signal-to-noise ratio; we have effectively masked
out the entire central region of NGC~3310, leaving only the outermost disc.   
Furthermore, most of the outer debris shown in Fig. 1 lies outside of 
the $D_{25}$ radius of 92\arcsec, and will not affect the final colour.
The B-band image and its pixel mask are shown in Figs. 5a and 5b.   When
calculated using only pixels within $D_{25}$ and not eliminated in the masks, we find a global
disc colour of $B-V = 0.44$, which is still bluer than the expected value
from \citet{deJong96}, but is rather more typical of an Sd galaxy.  It is, however, in
excellent agreement with our average colour for the outer radii of the surface brightness
profiles.  This suggests that $B-V \sim 0.4$ well represents the outer regions of the 
disc of NGC~3310.  While debris within $D_{25}$ still adds to the total colour of the
disc, its contribution is small since the debris is expected to be fainter than the galaxy
itself.  Furthermore, the photometry of Table 1 suggests that the debris in this system
would bias the galaxy toward a redder $B-V$.  Given the blue nature of the $B-V$ resulting
from the pixel masks, it is unlikely that overlying tidal debris is a significant 
contributor to the global galactic colour. 

%
%
\begin{figure}
\includegraphics[angle=-90.0,width=3.15in]{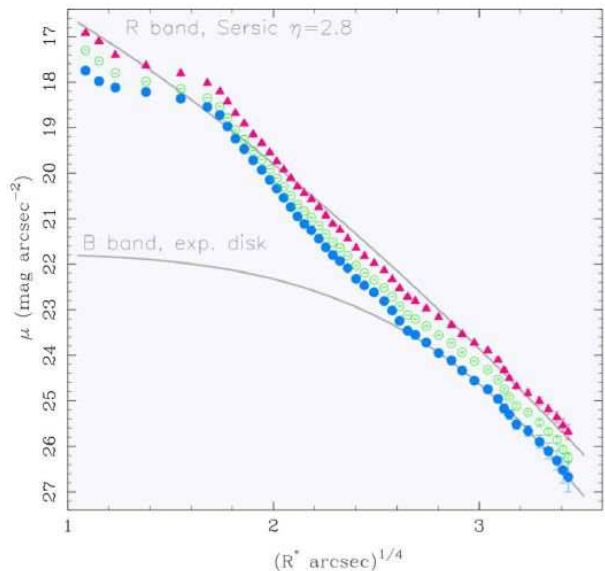}
\caption{Surface brightness profiles in $BVR$, plotted versus $R^{1/4}$.  The lower grey curve is the fit from Fig. 3, while the upper grey line shows a S\'ersic fit to the $R$-band profile.  This fit yields $1/\eta = 1/2.8$, rather than the $1/\eta = 1/4$ expected for a de Vaucouleurs profile.   }
\label{f4}
\end{figure}

%
%
\begin{figure*}
\begin{minipage}{6.5in}
\begin{tabular}{cc}
\includegraphics[width=3.15in]{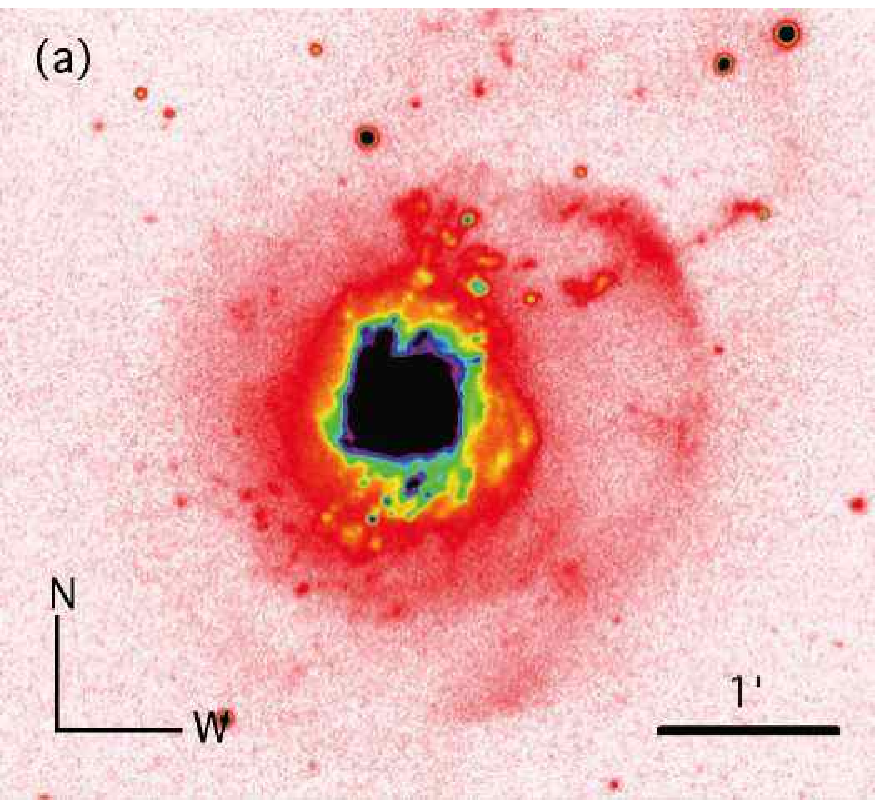} & 
\includegraphics[width=3.15in]{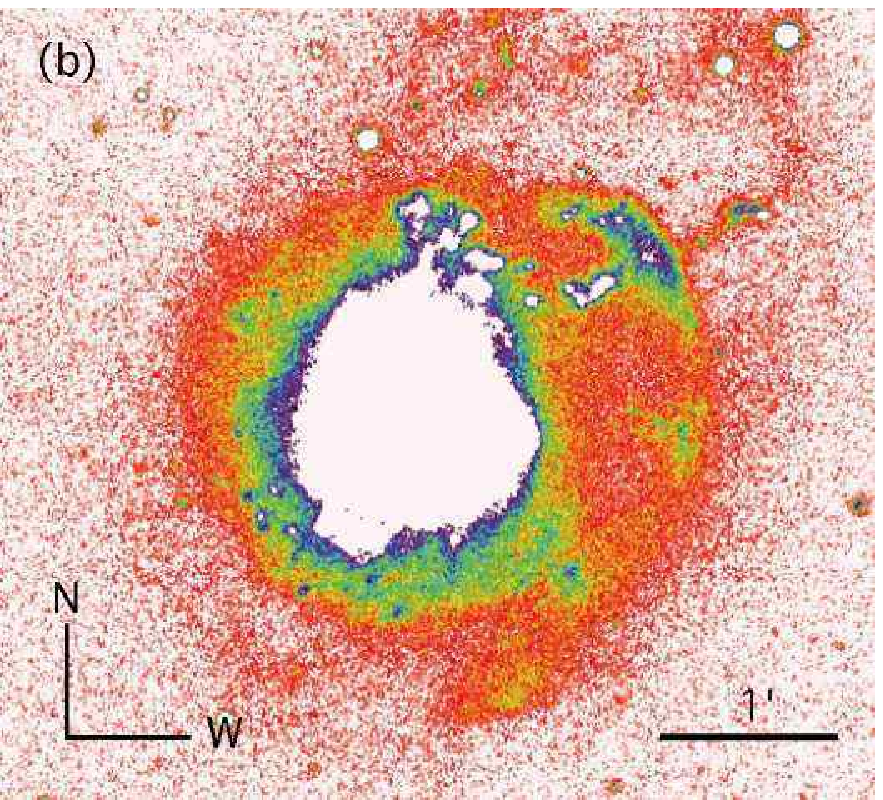} \\
\end{tabular}
\caption{ \textbf{(a)}  $B$-Band image of NGC~3310 taken with the WIYN 0.9m telescope on Kitt
Peak.  \textbf{(b)}  Same $B$-Band image of NGC~3310 with all pixels great than 5$\sigma$ above
the sky background removed.  This mask was used in all bands to estimate the
colour of the outer disc.  }
\label{f5}
\end{minipage}
\end{figure*}

\subsection{SparsePak Preliminary Results}

For each fibre in which H$\alpha$ was detected, we fit the line with a Gaussian
profile to find the peak of the line and measure its velocity.  Fibres that contained 
lines with $S/N > 3$ are shown in Fig. 6.  All but two lines were well fit by a single 
Gaussian profile.  For the two remaining fibres, which exhibited a double-peaked structure at the very top of the line, the profile was fit as a single line to obtain an average velocity.  In both cases, each actual peak's velocity deviates from the average fit by less than 3 per cent.  

Since the SparsePak data traces the warm ionised gas component in NGC~3310, these 
data suggest that NGC~3310 has maintained its disc structure and rotation in 
the star-forming component of its disc despite recent substantial mergers with this system.
While the old stars may have experienced a substantial disturbance, 
the young stars, as represented
by the H$\alpha$ emission shown in Fig. 6, appear to exist in a rotating system.   
The velocities shown in Fig. 6 are not corrected for inclination.  From $HI$ data,
\citet{mulder95} find $i = 52\degr$.  Earlier work focusing on H$\alpha$ emission finds
a slightly lower inclination of $i = 32\deg$.  While our observed $\Delta v = 100$
km~s$^{-1}$ across NGC 3310's disc is at first glance somewhat low for a typical disc
galaxy, when corrected for inclination (adopting the $HI$ value) this $\Delta v$
becomes 124 km~s$^{-1}$, yielding a rotational velocity of 62 km~s$^{-1}$.  While the
$HI$ exhibits a higher total velocity, the $HI$ data extend well beyond the range of
our SparsePak data.  At a radius of 40\arcsec, or 2.7 Kpc, our value is consistent with that
of \citet{mulder95}.  It is also interesting to note that our SparsePak data show an
initial increase in the velocity of the H$\alpha$ emitting gas, and then drops at
higher radii to values consistent with those from $HI$.  This trend is also noted in
\citet{vdk76,grc82}
The fact that the warm ionised gas is fairly smoothly rotating may provide some 
support for the idea that the disc in this system may be 
in the process of reforming in the manner described by \citet{fhg96}
after a gas-rich merger \citep{sh05}.  Or, rather NGC~3310 may have
maintained its disc structure despite multiple collisions.  If this is indeed the case, 
galactic discs may be surprisingly robust.

%
%
\begin{figure}
\begin{center}
\includegraphics[width=3.15in]{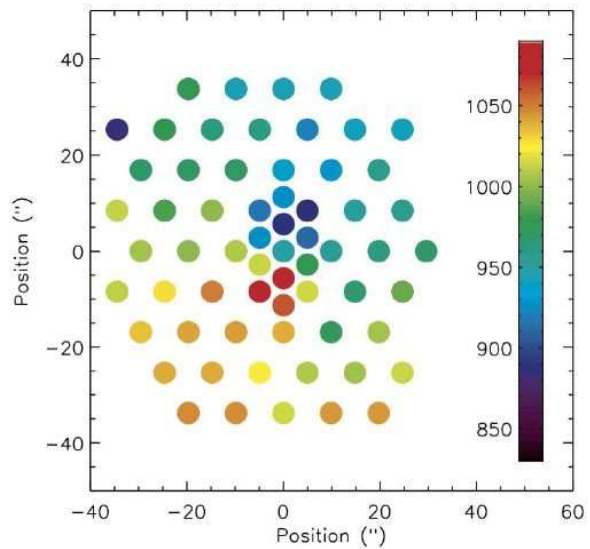}
\caption{ Radial velocity plot of NGC~3310.  Velocities were measured from the H$\alpha$ line in each fibre of the SparsePak array, and only those lines with $S/N > 3$ are shown.}
\label{f6}
\end{center}
\end{figure}

\subsection{Discussion}
\label{subsecdebris}

The $B-V$ calculated for the outer disc regions is redder than the global $B-V$
given in the RC3 which includes the bright star-forming regions.  However, 
$B-V$ for the debris regions ranges from 0.63 in region F to the South to
0.22 for the bright star-forming region, K, in the arrow.  That several regions are still
significantly redder than the underlying disc stars suggests that they did not originate in the
disc of NGC~3310.  One possibility is that they consist of the broken remains of an incoming,
disrupted companion galaxy - as has been suggested for a similar debris loop around NGC 5907 \citep{zheng99,shang98,sackett94,mbh94}.  
The bluer regions may reflect a subsequent 
(and currently fading) star formation episode among the debris.  

We also note the presence of an interesting colour gradient in NGC~3310's debris.  
In Figs. 7a and 7b, we show colour plots of NGC~3310 in $B-V$ for two different colour
ranges.  In these images, a colour gradient is apparent from regions F to E to D to C, 
changing from $B-V = 0.63$ in region F to $B-V = 0.42$ in region C.  
This could be caused by a propagating wave associated with an incoming companion
inducing star formation along the outer debris in the arc defined by F-C.  After a burst
of star formation, the $U-B$ colour will fade faster than $B-V$, which is less sensitive than
the $U$-band to the presence of the most massive stars.  Thus, the increasingly blue $B-V$, 
yet fairly constant $U-B$ from F to D, may represent a fading burst passing from F to D.
In this case, the arrow structure may represent the front edge of a propagating star-forming wave, 
leaving regions of recently compressed gas and new star formation in its wake.  Likewise, the star
formation rate in these red regions (F, J) might have been truncated some time ago
if they consist of material expelled as tidal structures from within NGC~3310. In
models of stellar populations that have had their star formation completely truncated, it
takes about $2 \times 10^8$ years to drive $B-V$ from 0.42 to 0.62 after completely stopping star
formation.  This may also explain the colour gradient in regions FEDC.  

It remains a possibility that the masks were simply insufficient 
to remove all the light from the starburst.  In this case, the pre-burst
disc stars may be redder, and the shells could have originated in the disc.  We further explore
the origins of these regions by comparison with spectral synthesis models in the following section.

%
%
\begin{figure*}
\begin{minipage}[b]{180mm}
\begin{tabular}{cc}
\includegraphics{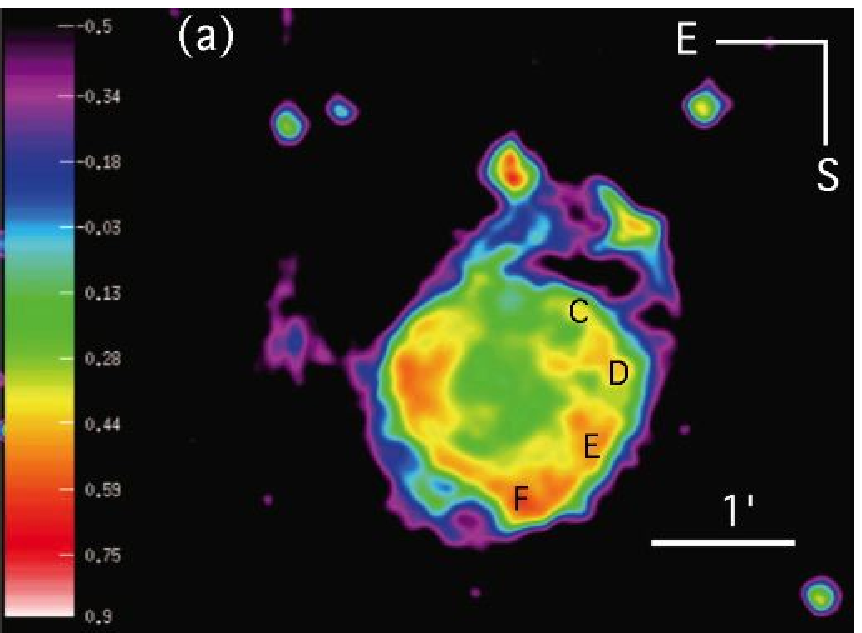} &
\includegraphics{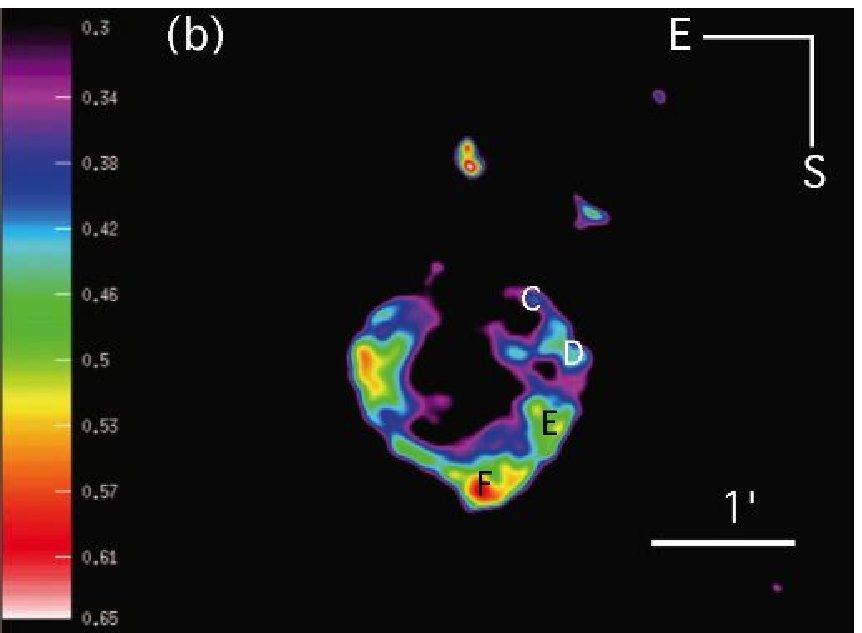} \\
\end{tabular}
\caption{B-V colour map of NGC~3310.  B- and V-Band images were smoothed using a
Gaussian profile to match their PSFs, calibrated photometrically and then subtracted.
The image on the right shows a smaller range in colour than the $B-V$ range shown in the left.   
The locations of regions F, E, D, and C are shown in each image.\hfill  }
\label{f7}
\end{minipage} 
\end{figure*}

\section{Comparison with GALEV Models}
\label{secgalev}

\subsection{Model Assembly}
\label{subsecmodels}

The GALEV stellar population synthesis models \citep{schulz02, anders03} 
involve two different components - one of which
represents the underlying, pre-burst galaxy, and another which represents the
starburst.  We can then compute colours for the 
combined system of disc and starburst.  GALEV provides two options for the underlying
disc component.  The first is a galactic disc that has experienced
constant star formation (for which we chose 2 M$_{\odot}$ yr$^{-1}$) 
throughout its lifetime.  This model, when combined with a
starburst model, defines the blue limit for a galaxy.  The other option is
to represent the underlying galaxy with an exponentially declining star formation
rate.  This model begins with a high rate of star formation that declines on a timescale
of $10^9$ years.  In this model, the galaxy is dominated by passive evolution and represents
the red limit for a galactic disc.  Each of these models have been calculated with
five different metallicities.  For comparison with our work, we have chosen an
intermediate metallicity value of $Z = 0.008$, since while NGC~3310 is thought to
have greater than solar metallicity in the nucleus, results from \citet{pastoriza93} suggests
that material outside the starburst ring is subsolar in metallicity.

The output for each model is a set of luminosities in different filters
for a given stellar mass for each time step from $t=0$ to beyond a Hubble time.  We 
combined the two models using:

\begin{equation}
L_{\lambda,tot} = L_{\lambda,ug} + b  L_{\lambda,SB} 
\end{equation}

where $L_{\lambda,ug}$ is the luminosity of the underlying galaxy component, per
solar mass at a given wavelength, $L_{\lambda,SB}$ is the luminosity per solar
mass of the overlying burst, and 
b is the strength of the burst by mass fraction.  For
$L_{\lambda,ug}$ we chose the luminosity at 12 Gyr to represent an existing disc
galaxy.  The luminosities (and colours) of the underlying galaxy are actually quite 
insensitive to age after approximately 5-6 Gyr, and thus colour is mostly dependent on 
which model is chosen for the underlying galaxy.  

The total luminosities were found for each band and
were converted to magnitudes and then colours.  These
colours form the evolutionary tracks seen in Fig. 8.

%
%
\begin{figure*}
\begin{tabular}{cc}
\includegraphics[width=3.15in]{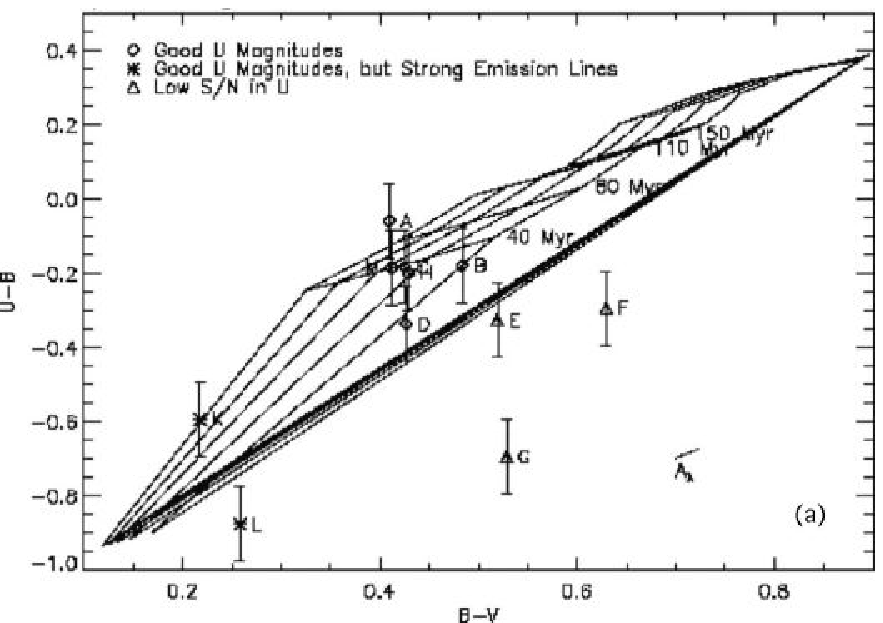} &
\includegraphics[width=3.15in]{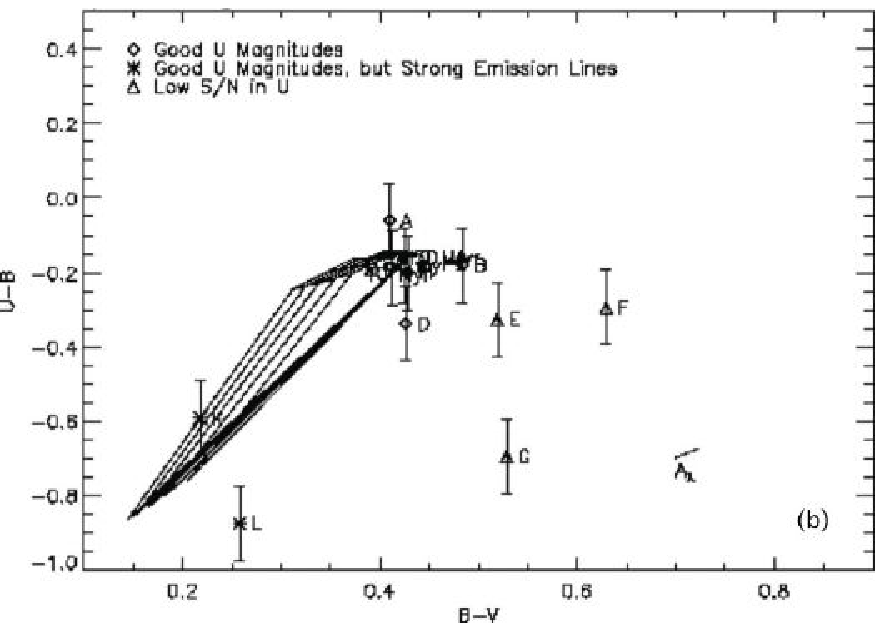} \\
\includegraphics[width=3.15in]{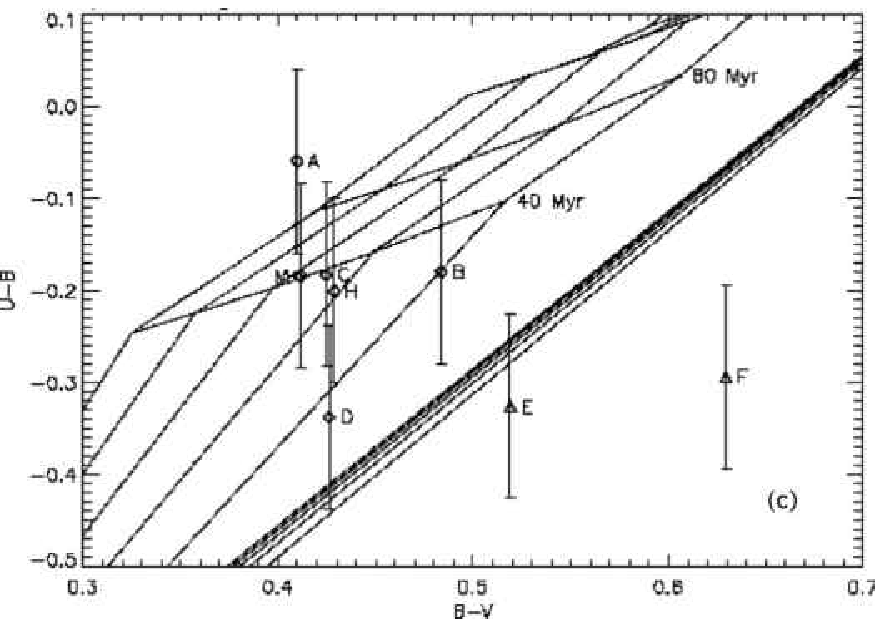}  & 
\includegraphics[width=3.15in]{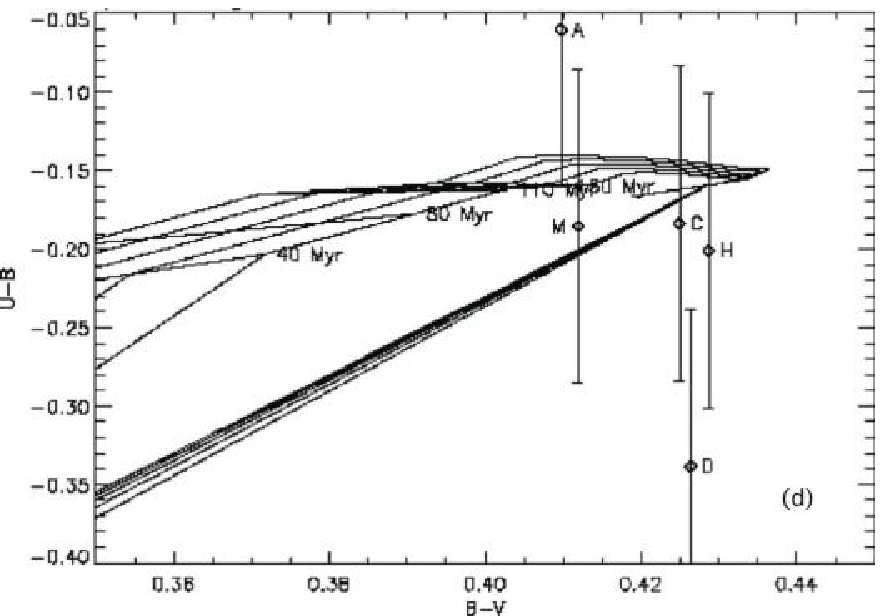} \\
\end{tabular}
\label{f8}
\caption{ \emph{Left:} \textbf{(a)} Lines show the models of exponential star formation with various burst models
overlaid.  The burst models range in strength from 0.08 to 2 per cent of the local stellar population
by mass.  Each of these bursts is represented by a curve, while lines of constant time
connect the burst models.   The figure below \textbf{(c)} shows an enlarged subset of the top figure.
\emph{Right:} \textbf{(b)} Same as the left, except underlying galaxy is represented by a constantly star-forming disc galaxy.  This disc has a colour of $B-V = 0.44$, chosen to be similar to the average colour of the outer regions of NGC~3310. The bottom image \textbf{(d)} shows an enlarged subset of the figure directly above.}
\end{figure*}

\subsection{Model Results}
\label{galevresults}

In order to explore the stellar populations resident in each debris region of NGC~3310,
we compared our photometry of the regions displayed in Fig. 1 with various 
models from the GALEV spectral synthesis models.  In Fig. 8, we show our
photometric results plotted over several model curves created using the GALEV
models.  

Each curve in Figs. 8a and 8c represents an underlying galaxy with exponentially declining
star formation that has a starburst overlaid.  The burst is represented as an
instantaneous event that takes the underlying, ancient galaxy (t = 12 Gyr) from
the reddest point in $U-B$ and $B-V$ space, to the bluest.  The colour then
slowly grows redder as the stars which were formed in the burst then age.  
Each curve represents a different strength of burst, ranging from 
0.8 to 2 per cent of the local stellar population by mass, and whose light then 
adds to that of the ancient stellar population.  

While the error bars are most likely higher in the $U$-band due to greater systematic 
errors, there are still trends evident in Fig. 8a.  Regions A, B, C, H, K and M 
are all consistent with burst strength of
0.014, i.e. a starburst involving 1.4 per cent of the local stellar population's mass.  They
are also consistent with a post-burst age of $50 \pm 20$ Myr.  While
region D is slightly bluer in $U-B$, it is also consistent with the rest in age,
although it requires a slightly lower burst strength, $b = 0.007$ i.e. with
only 0.7 per cent of that region's total mass.  It is possible that region D has
simply experienced less star formation than the aforementioned regions.  These 
results are consistent with the burst age found from NGC~3310 star clusters by \citet{deGrijs03b}.

While region L is lower than all the models, it is still consistent with the
rest of the points when the errors of the models themselves are considered ($\pm
0.1$ magnitudes).

Figs. 8b and 8d show the results for an underlying galaxy with a constant star formation rate plus
a starburst.  Note that the initial $B-V = 0.44$ for the constant star-forming model at $t=12$
Gyr. This is the same value of $B-V$ that we measure when we removed the starburst component from the
disc light using pixel masks.  If these masks successfully removed the starburst
contribution to the total light, then NGC~3310 should be well represented by a constantly star-forming disc
before the most recent burst occurred.  

In comparing our photometry to the models in Fig. 8b, we find that while a few of the 
bluer regions, such as K, A, M, C, and H overlap with these
evolutionary tracks, the rest of the debris network is inconsistent with these models, even
when the model errors of $\pm 0.1$ are taken into account.  While it is possible these regions
have a completely distinct evolutionary history from the rest of the debris, it is more likely
that their evolution is connected with the rest of the debris system, and that their bluer colours
indicate higher burst strengths in these regions, as is suggested by Fig. 8a.  We therefore
conclude that the exponentially decaying star formation rate plus burst model gives the better description for
these data.

\section{Summary}
\label{secsummary}

We have explored the photometric properties and the origins of NGC~3310's tidal debris.
This galaxy is currently undergoing a starburst induced by a complex merger.
We find a global $B-V = 0.30$ for NGC~3310 which is
consistent with \citet{deVauc91}.  In an attempt to separate the
contributions of the starburst and the underlying disc in the NGC~3310 system, we
created a mask in the B-band to eliminate all luminous blue regions in the disc.
This resulted in a global $B-V = 0.44$ for areas with little active star formation.  This
is also consistent with the average colour ($B-V$ = 0.42) calculated from 
the surface brightness profiles at the outer radii of NGC~3310.  
However, the colours of the shell and debris structures range in $B-V$ from as red 
as 0.63 to as blue as 0.22, as measured for region K in the star-forming arrow.  This suggests
that the debris did not originate in the disc.

Another possibility is we were unable
to completely separate the contribution of the bright starburst and the underlying 
disc, and that NGC~3310's underlying disc is truly redder than $B-V = 0.44$.  Since NGC~
3310 is UV bright any little remaining contamination from the burst could skew the results
to a bluer $B-V$.  In order to explore this, we assumed a red, underlying disc
comprised of an aged stellar population experiencing exponentially declining star
formation and overlaid starburst models of various strengths to create composite
colours.  We also created another set of composite colours
using a bluer underlying disc model with a modest yet constant star formation.
The model for the underlying galaxy in this latter case also matched our
calculated underlying disc colours of $B-V = 0.44$.  

The comparison of colours of debris structures with predictions of GALEV stellar
population synthesis models also suggests that these structures are not simply remains of NGC~3310's 
pre-merger stellar disc.  They are more likely the remains of a redder, disrupted companion
galaxy that was destroyed in merging with NGC~3310.  
In comparing the $B-V$ colours of regions in the debris with those of models, we find that they are
consistent with an ancient stellar population that has experienced a weak burst
of star formation in the past 50 Myr.  In general, the distinctive outer features of NGC~3310 have colours that imply  significantly different evolutionary histories than those of its main disc, and are
not consistent with the bluer constantly star-forming disc $+$ starburst models. 

The extensive debris network - specifically the large stellar loop, the two
extended HI tails and the stellar shells - as well as the blue underlying disc colours, the high gas content and the starburst may be indicative that we are witnessing the 
result of a major merger or multiple, smaller, simultaneous mergers and that the disc of
NGC~3310 is being reformed as suggested by \citet{ks01}.  

Our deep surface photometry indicates that the SBP of NGC~3310 
is best approximated by a S\'ersic model with an exponent 
$1/\eta = 1/2.8$, slightly larger that the value $1/\eta = 1/4$ that is expected for 
the de Vaucouleurs law. However, systematic residuals between the S\'ersic
fit and the observed SBPs suggest that a single fitting law cannot provide 
a perfect fit to the SBPs.
This fact, in connection with the exponential slope observed in all bands for 
$R > 60\arcsec$ suggest that a disc+starburst model gives a better approximation
to NGC3310's total light and supports the idea that the disc has survived the 
recent mergers experienced by NGC~3310.
This does not entirely rule out a major merger, however,
as \citet{sh05} suggest that if disc galaxies are sufficiently gas rich, then discs can 
survive equal mass mergers, implying that discs may be surprisingly robust.  
However, since there is some evidence of older stellar populations 
in this system, it is perhaps more likely that we are seeing the growth of NGC~3310's 
disc through the accretion of multiple, small companions onto a preexisting disc.   

Most of the debris is consistent with a small burst of star formation 
in the past 50 Myr, (which is consistent with the ages of the young
stellar clusters in the central starbursting region \citep{deGrijs03a}), and 
recent work by \citet{knapp06} suggests that a secondary star formation event, 
as indicated by the presence of a young stellar cluster, may have occurred in this 
system approximately 750 Myr ago.  This discrepancy may result from either 
unique, sequential mergers 
with NGC~3310, or the incoming companion and its resulting debris completing multiple orbits 
before finally being destroyed.  Either case may help account for the observed range of colours 
in the outer debris network.

Future 2D integrated field unit spectroscopy with improved velocity resolution 
would be useful for further
examining the stability of the disc and exploring the likelihood that we are
witnessing a reforming disc.  It may also be useful to obtain kinematic
information on the shells, although their low optical surface brightness and low
HI content will make it difficult to obtain reasonable signal-to-noise ratios in
these regions.  In lieu of such data, dynamical modelling of this system
would provide further insight into the origins of the surrounding stellar debris. 

\section*{Acknowledgements}

EHW and JSG thank the National Science Foundation for support through grant
NSF AST98-03018 to the University of Wisconsin and the University of
Wisconsin-Madison Graduate School for additional funding.
EHW would also like to thank the National Space Grant College and Fellowship Program 
and the Wisconsin Space Grant Consortium for their generous funding of
this project as well as Bill Harris for his helpful discussions.  We would also like
to thank the referee for their careful review and insightful comments.

This research has made use of the NASA/IPAC Extragalactic Database (NED) which is operated by the Jet Propulsion Laboratory, California Institute of Technology, under contract with the National Aeronautics and Space Administration.

\bsp

\label{lastpage}

\end{document}